\documentstyle[aps,twocolumn,epsf]{revtex}

\begin{document} 

\title{Modified Scaling Relation for the Random-Field Ising Model}
\author{U.~Nowak, K.~D.~Usadel, and J.~Esser}
\address{
  Theoretische Tieftemperaturphysik, Gerhard-Mercator-Universit\"{a}t-Duisburg,
  47048 Duisburg/ Germany\\
  e-mail: uli@thp.uni-duisburg.de 
}

\date{\today}
\maketitle

\begin{abstract}
  We investigate the low-temperature critical behavior of the three
  dimensional ran\-dom-field Ising ferromagnet. By a scaling analysis
  we find that in the limit of temperature $T \rightarrow 0$ the usual
  scaling relations have to be modified as far as the exponent
  $\alpha$ of the specific heat is concerned. At zero temperature, the
  Rushbrooke equation is modified to $\alpha + 2 \beta + \gamma = 1$,
  an equation which we expect to be valid also for other systems with
  similar critical behavior.  We test the scaling theory numerically
  for the three dimensional random field Ising system with Gaussian
  probability distribution of the random fields by a combination of
  calculations of exact ground states with an integer optimization
  algorithm and Monte Carlo methods. By a finite size scaling analysis
  we calculate the critical exponents $\nu \approx 1.0$, $\beta
  \approx 0.05$, $\bar{\gamma} \approx 2.9$ $\gamma \approx 1.5$ and
  $\alpha \approx -0.55$.
\end{abstract}



Above two dimensions, the ferromagnetic random-field Ising model has an
ordered phase for low temperatures and small random-fields as was
proven by Imbrie \cite{imbrie} and later also by Bricmont and
Kupiainen \cite{bricmont}. For larger fields the system develops a
domain state \cite{imry} which has been shown to have a complex and
fractal structure \cite{esser}. It is now widely believed that the
phase transition from the ordered to the disordered phase is of second
order. In three dimensions, the values of some of the critical
exponents are now well established, like $\beta \approx 0.05$ and $\nu
\approx 1$. Although real space renormalization yields deviating
results concerning $\nu$ (see e.~g. \cite{falicov}) the values for
$\beta/\nu$ are in the same range. However, a complete set of values
of the critical exponents fulfilling the predicted set of scaling
relations \cite{bray,nattermann,villain} could not be established,
neither by experimental measurements (for a review, see
\cite{belanger}) performed usually on diluted antiferromagnets in a
magnetic fields which are thought to be in the same universality class
\cite{fishman} nor by any numerical methods (see e.~g.~
\cite{rieger}). Especially, the value of $\alpha$ - and even its sign
- is highly controversial.

The Hamiltonian of the RFIM in units of the nearest neighbor coupling
constant $J$ is 
\begin{equation}
H=-\sum_{<ij>} \sigma_i \sigma_j - \sum_i B_i \sigma_i.
\end{equation}
The first sum is over the nearest neighbors and the spin variables
$\sigma_i$ are $\pm 1$. The random-fields $B_i$ are taken from a
Gaussian probability distribution $P(B_i) \sim \exp(-(B_i/\Delta)^2/2)$.

We assume that there is a zero temperature fixed point at a finite
value $\Delta_c$ of the random-field width. Introducing the scaling
variable $f = \Delta_0 -\Delta -g(T)$ (see also \cite{nattermann})
where the condition $f = 0$ describes the critical line we
expect the same critical behavior no matter if we vary the temperature
or the random-field strength. Hence, for the singular part of the internal
energy $E$ it should be
\begin{equation}
  c \sim \frac{\partial E}{\partial T}
    \sim \frac{\partial E}{\partial \Delta}
    \sim |f|^{-\alpha},
\end{equation}  
and for the singular part of the free energy $F$ 
\begin{equation}
\label{e:c}
  c \sim -T \frac{\partial^2 F}{\partial T^2}
    \sim \frac{\partial^2 F}{\partial \Delta^2}
    \sim |f|^{-\alpha}.
\end{equation}  
In the limit of low temperatures $F$ equals $E$ and the question
arises which derivative - first or second - with respect to the
random-field strength yields the exponent of the specific heat?

The scaling ansatz for the singular part of the free energy for
zero homogenous magnetic field is:
\begin{equation}
  F(T,f) = |f|^{1/x_2} {\cal F}^{\pm} \left( \frac{T}{|f|^{x_1/x_2}}\right) 
\end{equation}
Hence, for the most singular part of the specific heat it is $c \sim
|f|^{1/x_2-2}$ and consequently $\alpha = 2-1/x_2$, as usual.  On the
other hand, for fixed critical random-field $\Delta = \Delta_0$ and in
the limit of temperature $T\rightarrow 0$ the prefactor $T$ in 
Eq.~\ref{e:c} becomes critical and hence it is $\alpha = 1-1/x_2$.
This is also consistent with the scaling behavior of $\frac{\partial
  E}{\partial \Delta} = \frac{\partial }{\partial \Delta} (F+TS)$ with
$S=-\frac{\partial F}{\partial T}$ which follows from the scaling
ansatz for the free energy. The most relevant singular terms are:

\begin{eqnarray}
\label{e:dedh}
  \frac{\partial E}{\partial \Delta} &=&
        - T |f|^{1/x_2-2} \frac{\partial |f|}{\partial \Delta}
        \frac{\partial |f|}{\partial T} \frac{1-x_2}{x_2^2}
        {\cal F}^{\pm} \nonumber \\
&&        + |f|^{1/x_2-1} \frac{\partial |f|}{\partial \Delta}
        \frac{1}{x_2} {\cal F}^{\pm}
        + \ldots
\end{eqnarray}
In the limit $T \rightarrow 0$ the first term vanishes and only the
second term is observed in the specific heat leading to $\alpha =
1-1/x_2$ as above. Consequently, it follows by standard scaling theory
that the scaling relations in the limit of temperature $T \rightarrow
0$ have to be modified with respect to $\alpha$. For zero temperature,
the equation corresponding to the Rushbrooke equation has the form
\begin{equation}
  \label{e:rush}
  2 \beta + \gamma = 1-\alpha.
\end{equation}
Additionally, it is remarkable that in Eq. \ref{e:dedh} the second more
singular term is small for either small $T$ or small $\partial |f|/
\partial T$. Therefore, one can expect to observe the anomalous zero
temperature critical behavior as long as the critical line is flat.

A similar, although more complicated consideration holds for the
specific heat. Building the derivative $c = -T \frac{\partial^2
  F}{\partial T^2}$, the most relevant singular terms are:

\begin{eqnarray}
  c &=&
    -T |f|^{1/x_2-2} \frac{1-x_2}{x_2^2} (\frac{\partial |f|}{\partial T})^2
        {\cal F}^{\pm} \nonumber \\
&&      -T |f|^{1/x_2-1} \frac{1}{x_2} \frac{\partial^2 |f|}{\partial T^2}
        {\cal F}^{\pm}
      + \ldots 
\end{eqnarray}
The most relevant term has the square of the slope $\frac{\partial
  |f|}{\partial T}$ as a prefactor while the next relevant term has
the curvature $\frac{\partial^2 |f|}{\partial T^2}$ as a prefactor. If
the critical line starts horizontal at $T=0$ but with a finite
curvature the most relevant term will be suppressed and the unusual
less critical behavior will be observed, yielding $\alpha = 1-1/x_2$
for low temperatures. Note that in both cases, for $c$ as well as for
$\frac{\partial E}{\partial \Delta}$ the $T \rightarrow 0$ critical
behavior is an inflection point since $\frac{\partial |f|}{\partial
  \Delta}$ and $\frac{\partial^2 |f|}{\partial T^2}$ change the sign
at the critical point.  However, for finite temperatures close to the
critical point a crossover to the "normal" critical behavior can be
expected.

In order to test these arguments numerically we consider the three
dimensional RFIM and calculate exact ground states (EGS) using an
optimization algorithm well known in graph-theory. The Ising-system is
mapped on an equivalent transport network, and the maximum flow is
calculated using the Ford-Fulkerson algorithm
\cite{ford,picard,ogielski}. We used a simple cubic lattice with
periodical boundary conditions and linear lattice sizes varying from
$L=6$ to $L=20$.  From the spin configurations of the ground state, we
can calculate the magnetization $M = \left[ m \right]_{\mbox{av}}$,
where $m = \frac{1}{L^3}\sum_i \sigma_i$, the internal energy, $E =
\left[h \right]_{\mbox{av}}$ where $h = \frac{1}{L^3} H$ and the
disconnected susceptibility $\chi_{\mbox{dis}} = L^3
\left[m^2\right]_{\mbox{av}}$, where the square brackets denote an
average taken over 30-1900 random-field configurations, depending on
the system size.  The advantage of the numerical technique above is
that it supplies equilibrium information.  But on the other hand it is
restricted to zero temperature. Therefore, we combine the EGS
calculation with Monte Carlo (MC) methods. Starting at zero temperature
with an EGS-spin configuration for a certain set of random-fields we
heat the system slowly using the standard heat bath algorithm. Hence,
we get MC data at low temperatures, $T < T_c/2$, which are close to
equilibrium. We checked that by heating the system with decreasing
heating rates until no further change of the data was visible.  Using
this MC method we can additionally calculate the susceptibility $\chi
= \frac{L^3}{T} \left[\langle m^2 \rangle -\langle m \rangle^2
\right]_{\mbox{av}}$, and the specific heat $c = \frac{L^3}{T^2}
\left[\langle h^2 \rangle -\langle h \rangle^2 \right]_{\mbox{av}}$,
where the angles denote a thermal average.

From the scaling relations above follow the finite size scaling relations

\begin{equation}
  M = L^{-\beta/\nu} \tilde{M}\left((\Delta-\Delta_c)L^{1/\nu}\right)
\end{equation}
for the magnetization and
\begin{equation}
  \chi_{\mbox{dis}} = L^{\bar{\gamma}/\nu}
               \tilde{\chi}\left((\Delta-\Delta_c)L^{1/\nu}\right)
\end{equation}
for the disconnected susceptibility.

\begin{figure}
  \begin{center}
    \epsfxsize=8cm
    \epsffile{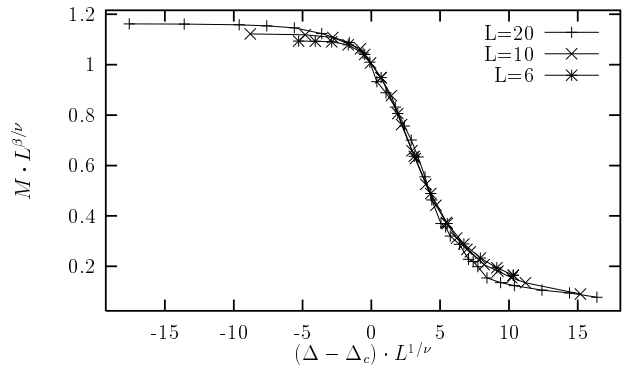}
  \end{center}
  \caption{Scaling plot of the magnetisation from EGS.}
  \label{f:fs_mag}
\end{figure}

Figure \ref{f:fs_mag} shows the scaling plot for the magnetization
data from EGS calculations as described above yielding $\Delta_0 =
\Delta_c(T=0) = 2.37 \pm 0.05$, $\nu = 1.0 \pm 0.1$, and $\beta = 0.05
\pm 0.05$.  These are values which are not surprising and in agreement
with most of the previous work, especially the previous EGS
calculations of Ogielski \cite{ogielski}. The error-bars are estimated
since there is no straight-forward way to extract error-bars from a
finite-size scaling plot.

Using the same values for $\Delta_c$ and $\nu$ as above from the
scaling plot for the disconnected susceptibility (not shown) we get
$\bar{\gamma} = 2.9 \pm 0.3$ which is also in agreement with most of
the previous work.

\begin{figure}
  \begin{center}
    \epsfxsize=8cm
    \epsffile{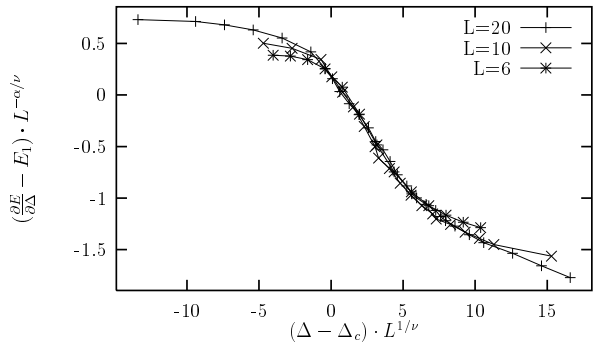}
  \end{center}
  \caption{Scaling plot of the derivative of energy from EGS.}
  \label{f:fs_ener}
\end{figure}

The first derivative of the ground state energy shows the critical
behavior of the specific heat. The behavior of $E$ can be understood
by a series expansion of the energy in the vicinity of the critical
point:
\begin{equation}
  E(\Delta) = E_0 + E_1(\Delta-\Delta_c) + E_s
  (\Delta-\Delta_c)^{1-\alpha} + \ldots
\end{equation}
The $E_1$-term is important since as we argued above $\alpha$ can be
expected to be negative for low temperatures following Eq.
\ref{e:rush}.  Hence, the finite size scaling form is
\begin{equation}
  \frac{\partial E}{\partial \Delta} - E_1 = L^{\alpha/\nu}
  \tilde{E}\left((\Delta-\Delta_c)L^{1/\nu}\right)
\end{equation}
for the derivative of $E$ in the critical region.  Differentiating our
energy data numerically, we obtained the scaling plot shown in Figure
\ref{f:fs_ener}. Once more we used the same values for $\Delta_c$ and
$\nu$ as above and chose $E_1$ such that $\frac{\partial E}{\partial
  \Delta} = E_1$ at the inflection point. Note, that we neglect here a
possible size dependence of the analytic parts of the energy, i.~e. a
possible $L$-dependence of $E_1$ which is obviously very small as
Figure \ref{f:fs_ener} demonstrates.  This analysis leads to $\alpha =
-0.55 \pm 0.2$.

\begin{figure}
  \begin{center}
    \epsfxsize=8cm
    \epsffile{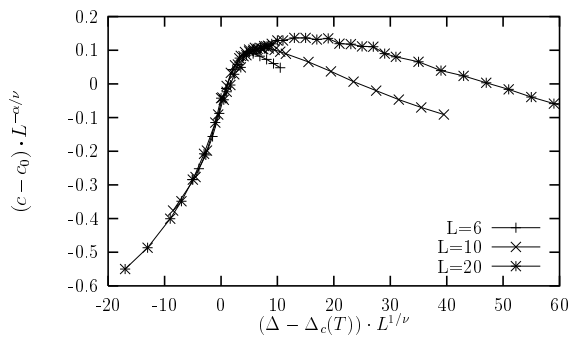}
  \end{center}
  \caption{Scaling plot of the specific heat from MC simulation. $T=1.4$.}
  \label{f:fs_c}
\end{figure}

The approach to determine $\alpha$ for finite temperatures is a direct
MC simulation of the specific heat. Figure \ref{f:fs_c} shows the
corresponding data. The ground state spin configurations were used as
initial spin configurations for a MC simulation - of course of systems
with identical random-field configuration. Then the systems were
slowly heated (10000 MCS per temperature with temperature steps of
0.2). Data are shown for $T = 1.4$ which is roughly 30\% of the
critical temperature at zero field. We do not find any divergence of
the specific heat, i.~e. no size dependence of the maximum of $c$.
Hence, as above we analyzed the data subtracting the value of the
energy at the inflection point $c_0$. We took the values
$\Delta_c(T=1.4) = 2.35$ and $\nu = 1$ from MC simulation data of $M$
for the same temperature (data not shown here). Our analysis yields
once more $\alpha = -0.55 \pm 0.2$. The value $\Delta_c(T=1.4)$ is
very close to the zero-temperature value $\Delta_0$, confirming that
the critical line is nearly horizontal in the low temperature region.
Hence, as discussed above the true critical behavior fulfilling the
Rushbrooke equation is hard to observe and within our numerical
accuracy we can only find the zero temperature exponent.

A standard finite size analysis of the MC data for $\chi$ yields
$\gamma = 1.5 \pm 0.2$ (not shown). This value is in agreement with
the previous work of Ogielski. Nevertheless there are deviations from
previous results from series expansion where $\gamma$ was found to be
significantly higher \cite{gofman}.  We analyzed MC data also for
other temperatures but we did not find any significant temperature
dependence of the critical exponents in the temperature range $0 < T <
2$ which is nearly one half of the phase diagram.

To summarize, the values we determined for the critical exponents of
the 3D RFIM are in good agreement with many of the previous works,
experimental as well as theoretical.  The exponents $\gamma$ and
$\bar{\gamma}$ fulfill the Schwartz-Soffer equation $\bar{\gamma} = 2
\gamma$ \cite{schwartz}.  The modified hyperscaling-relation
\cite{grinstein} which can be written in the form (without $\alpha$)
$\bar{\gamma} = D \nu -2 \beta$ is also fulfilled by our exponents.

 It is the aim of this work to calculate as many exponents as possible
independently in order to test Eq.\ref{e:rush} which is the most
important aspect of our work. It is derived from the standard scaling
ansatz for the zero temperature fixed point. Therefore we expect this
equation to be valid also for other systems for which the discussed
scaling ansatz is true.  Candidates may be the random-field-Heisenberg
model in appropriate dimensions. For the RFIM in higher dimensions new
results suggest that there is a break of universality (i.~e. the
critical behavior depends on the kind of the distribution of the
random-fields) as was shown in ref. \cite{swift} for four dimensions
and earlier in refs.  \cite{schneider} and \cite{aharony} for the mean
field solution of the RFIM. Additionally, it was shown \cite{mezard}
that there is replica symmetry breaking for the mean field solution of
a random field model with $m$-component-spins in the limit of large
$m$.  However, the replica symmetric solution of the RFIM with a
Gaussian distribution of random fields has a zero temperature fixed
point and indeed, in the limit $T \rightarrow 0$ the mean-field
exponents fulfill eq. \ref{e:rush} since it is $\beta = 0.5$, $\gamma
= 1$ and $\alpha = -1$ which can be inferred from a remark in ref.
\cite{schneider} stating that the entropy vanishes as $T \rightarrow
0$ linearly with $T$".

We argued that the crossover to the normal critical behavior might be
hard to observe as long as the critical line is horizontal.  We
directly determined the controversially discussed exponent $\alpha$
for zero temperature yielding $\alpha = -0.55$ confirming the validity
of Eq.\ref{e:rush}. The same value is also observed for finite but low
temperatures, the crossover to the true critical behavior cannot be
observed within our numerical accuracy.  Surprisingly,
this value is also in agreement with recent Monte Carlo simulations
\cite{rieger} as well as with recent experimental results
\cite{karszewski}. Both did not find a divergence of the specific heat
but $\alpha \le 0$, although these measurements and simulation,
respectively, were performed for higher temperatures where true
critical behavior should be easier to observe. 

{\bf Acknowledgments:} The authors thank A.~Hucht for providing his
program {\it fsscale} which simplified the finite size scaling work
enormously.

\end{document}